\documentclass[12pt]{iopart}

\usepackage{graphicx}
\usepackage{subfig}
\usepackage{amssymb}
\usepackage{amsthm} 
\usepackage{mathptmx}
\usepackage{nicefrac}
\usepackage{orcidlink}
\usepackage{xcolor}
\usepackage[switch]{lineno} 
\usepackage[T1]{fontenc}

\begin{document}

\title[]{First Kaonic Boron Isotopes Measurements with SIDDHARTA-2 at DA$\Phi$NE\footnote{We dedicate this work to the memory of Prof. C. Guaraldo and Prof. J. Zmeskal, whose contributions were essential to the success of the kaonic boron measurements. This work would not have been possible without them.}}

\author{
D Sirghi$^{1,2,3*}$, M Iliescu$^{2}$, F Sgaramella$^{2**}$, L Abbene$^{4,2}$, C Amsler$^{5***}$, F Artibani$^{2,6}$, M Bazzi$^{2}$, G Borghi$^{7,8}$, D Bosnar$^{9}$, M Bragadireanu$^{3}$, A Buttacavoli$^{4,2}$, M Carminati$^{7,8}$, A Clozza$^{2}$, F Clozza$^{2,10}$, L De Paolis$^{2}$, R Del Grande$^{11,2}$, K Dulski$^{2,12,13}$, C Fiorini$^{7,8}$, I Friščić$^{9}$, C Guaraldo$^{2,^\dagger}$, P Indelicato$^{14}$, M Iwasaki$^{15}$, A Khreptak$^{12,13,2}$, S Manti$^{2}$, J Marton$^{5,****}$, P Moskal$^{12,13}$, H Ohnishi$^{16}$, K Piscicchia$^{1,2}$, F Principato$^{4,2}$, A Scordo$^{2}$, M Silarski$^{12}$, F Sirghi$^{2,3}$, M Skurzok$^{12,13,2}$, A Spallone$^{2}$, K Toho$^{16,2}$, O Vazquez Doce$^{2}$, J Zmeskal$^{5,^\dagger}$, C Curceanu$^{2,3}$}

\address{$^{1}$ Centro Ricerche Enrico Fermi, Museo Storico della Fisica e Centro Studi e Ricerche "Enrico Fermi", Roma, Italy}
\address{$^{2}$ Laboratori Nazionali di Frascati INFN, Frascati, Italy}
\address{$^{3}$ IFIN-HH, Institutul National Pentru Fizica si Inginerie Nucleara Horia Hulubei, 30 Reactorului, 077125, Magurele, Romania}
\address{$^{4}$ Department of Physics and Chemistry (DiFC), Emilio Segrè, University of Palermo, Palermo, Italy}
\address{$^{5}$ Stefan Meyer Institute for Subatomic Physics, Vienna, Austria}
\address{$^{6}$ Università degli Studi di Roma Tre, Dipartimento di Fisica, Roma, Italy}
\address{$^{7}$ Politecnico di Milano, Dipartimento di Elettronica, Informazione e Bioingegneria, Milano, Italy}
\address{$^{8}$ INFN Sezione di Milano, Milano, Italy}
\address{$^{9}$ Department of Physics, Faculty of Science, University of Zagreb, Zagreb, Croatia}
\address{$^{10}$ Università degli Studi di Roma Tor Vergata, Dipartimento di Fisica, Roma, Italy}
\address{$^{11}$ Faculty of Nuclear Sciences and Physical Engineering, Czech Technical University in Prague, Břehovà 7, 115 19, Prague, Czech Republic}
\address{$^{12}$ Faculty of Physics, Astronomy, and Applied Computer Science, Jagiellonian University, Kraków, Poland}
\address{$^{13}$ Center for Theranostics, Jagiellonian University, Krakow, Poland}
\address{$^{14}$ Laboratoire Kastler Brossel, Sorbonne Université, CNRS, ENS-PSL Research University, Collège de France, Case 74; 4, place Jussieu, F-75005 Paris, France}
\address{$^{15}$ RIKEN, Tokyo, Japan}
\address{$^{16}$ Research Center for Accelerator and Radioisotope Science (RARiS), Tohoku University, Sendai, Japan}

\address{$^\dagger$ Deceased}
\ead{$^*$ Diana.Laura.Sirghi@lnf.infn.it (Corresponding Author)}
\ead{$^{**}$ francesco.sgaramella@lnf.infn.it (Corresponding Author)}
\ead{$^{***}$ Now at Marietta Blau Institute for Particle Physics, Vienna, Austria}
\ead{$^{****}$ Now at Atominstitut, Universitat Wien, 1020, Vienna, Austria }

\vspace{10pt}

\begin{abstract}

A precision measurement of X-ray transitions in kaonic boron, performed by the SIDDHARTA-2 collaboration at the DA$\Phi$NE collider, is reported. The energies and yields of the $5g\rightarrow4f$ and $4f\rightarrow3d$ transitions were determined for both boron isotopes, kaonic ${}^{10}$B and  kaonic ${}^{11}$B.

For the $5g\rightarrow4f$ transition, the measured energies are $7064.62 \pm 16.93~(\mathrm{stat.}) \pm 2.00~(\mathrm{sys.})$~eV for kaonic ${}^{11}$B and $6920.96 \pm 58.23~(\mathrm{stat.}) \pm 2.00~(\mathrm{sys.})$~eV for kaonic ${}^{10}$B. For the $4f\rightarrow3d$ transition, the corresponding values are $15293.33 \pm 4.80~(\mathrm{stat.}) \pm 5.30~(\mathrm{sys.})$~eV and $15180.11 \pm 20.86~(\mathrm{stat.}) \pm 5.30~(\mathrm{sys.})$~eV, respectively.

The yields for the $5g\rightarrow4f$ transition are $0.076 \pm 0.013~(\mathrm{stat.})^{+0.012}_{-0.011}~(\mathrm{sys.})$ for kaonic ${}^{11}$B and $0.079 \pm 0.014~(\mathrm{stat.})~^{+0.013}_{-0.011}~(\mathrm{sys.})$ for kaonic ${}^{10}$B. For the $4f\rightarrow3d$ transition, the corresponding yields are $0.115 \pm 0.006~(\mathrm{stat.})~^{+0.002}_{-0.005}~(\mathrm{sys.})$ and $0.107\pm 0.007~(\mathrm{stat.})~^{+0.002}_{-0.005}~(\mathrm{sys.})$, respectively.

No statistically significant deviation from pure electromagnetic (QED) calculations was observed in the  measurement of the $4f\rightarrow3d$ X-ray transition in kaonic ${}^{11}$B. Interpreted as upper limits, these results impose stringent constraints on the strong-interaction energy shift and  width of the 3d level  in light nuclei.  Translating these limits into bounds on phenomenological kaon–nucleus optical potentials, and, within specific theoretical models, on the complex scattering amplitude, we constrain and disfavor scenarios predicting large shifts or widths in boron. These results demonstrate, for the first time, the feasibility of precision studies of light kaonic atoms using solid targets, enabling investigations of kaon interactions with few-nucleon systems.

\end{abstract}

\vspace{2pc}
\section{Introduction}
\label{sec1:Introduction}
The strong interaction, described by Quantum Chromodynamics (QCD) within the Standard Model, governs the formation of hadrons from quarks and gluons, including nucleons. A comprehensive understanding of this interaction is essential for interpreting a broad range of physical phenomena, spanning from the subnuclear structure of matter to astrophysical processes such as the composition and equation of state of neutron stars \cite{DePietri:2019khb,Merafina:2020ffb,Akaishi:2016uvw,Drago:2019tbs}. While high-energy QCD can be treated perturbatively, the low-energy regime remains challenging due to its non-perturbative nature.
Hadronic atoms, in which a negatively charged hadron replaces an electron in an atomic orbit, offer a unique and precise tool to access the low-energy regime of the strong interaction. 
Among them, kaonic atoms play a particularly significant role because they contain a strange quark, providing a direct probe of QCD in the strangeness sector \cite{Curceanu:2020kkg}. The energy levels and X-ray transition spectra of kaonic atoms are strongly affected by the hadron–nucleon strong  interaction, allowing for direct experimental access to kaon–nucleon and kaon–nucleus interactions.  

The measurements performed at KEK, DA$\Phi$NE and J-PARC (such as  KpX, DEAR, SIDDHARTA experiments \cite{RevModPhys.91.025006}) have resolved the longstanding “kaonic hydrogen puzzle,” revealing a repulsive-type ground-state shift and providing stringent constraints on theoretical models of the low-energy $\bar{K}N$ interaction. With the advent of the SIDDHARTA-2 experiment at DA$\Phi$NE and the E62 experiment at J-PARC, a new era of precision spectroscopy of kaonic atoms has begun, leading to the first high-accuracy measurements of kaonic helium and kaonic deuterium \cite{Curceanu:2026zjg}. These results open new opportunities to investigate multi-nucleon dynamics and to test the strong interaction in its nonperturbative regime.

The next step toward a comprehensive understanding of the strong interaction in the strangeness sector is the detailed study of kaon interactions with few-nucleon systems. In the  case of kaonic deuterium, three-body dynamics can be addressed by solving Faddeev-type equations, which explicitly treat the coupled K$^{-}$NN system. However, this approach becomes intractable for heavier nuclei, where kaon–multi-nucleon interactions must be effectively modeled  and consistently incorporated to obtain reliable theoretical predictions.


Of particular importance are therefore the light kaonic atoms, where  K$^{-}$NN interactions are especially pronounced and strongly affect the atomic structure. Previous measurements for these systems suffer from large uncertainties: data for kaonic $^{6}$Li are still missing, X-ray yields for lithium (Z=3) and beryllium (Z=4) remain poorly determined, and no reliable information exists for boron (Z=5) \cite{Friedman:1994hx}. 
In the past, spectroscopy of light kaonic atoms was technically challenging due to the limited energy range of radiation detectors, resulting in an incomplete experimental database  affected by significant systematic uncertainties \cite{Friedman:1994hx}.

Among the light systems, kaonic $^{10}$B and $^{11}$B provide a rare and powerful opportunity to isolate isospin-dependent components of the kaon–nuclei interaction. The Coulomb potential and QED corrections are nearly identical in both systems. Any observed differences in level shifts or widths can therefore be directly  attributed to nuclear-structure effects, particularly the neutron number and its spatial distribution \cite{RevModPhys.91.025006,10.1093/ptep/ptae189}.

Using a new generation of Silicon Drift Detectors (SDDs) for X-ray spectroscopy, characterized by excellent energy and timing resolutions and capable of operating in the high-radiation environment of particle accelerators \cite{Sgaramella:2022rbl}, the SIDDHARTA-2 collaboration has performed high-precision measurements of kaonic-atom transitions at the DA$\Phi$NE collider \cite{Curceanu:2026zjg}. In this work, we report the first measurement of kaonic  $^{10}$B and $^{11}$B X-rays, specifically the atomic transitions $5g\rightarrow4f$ and $4f\rightarrow3d$. These results provide accurate values for the transition energies and yields, contributing to a deeper understanding of kaonic-atom cascade dynamics and paving the way for measurements aimed at precision QED tests.
The measurement also serves as a feasibility demonstration of the SIDDHARTA-2 apparatus technique for future studies of the kaonic boron $3d\rightarrow2p$ transition (the 2p level), where strong-interaction effects are expected to be large, as well as for measurements of other light kaonic atoms, such as kaonic lithium and beryllium, on both high- and low-n levels.

Moreover, the global analyses of kaonic atom data, such as those performed  in \cite{Batty1997,Friedman2007,Friedman2017} and within Chiral Effective Field Theory models\cite{Weise2010,Ikeda2011,Hyodo2012}, rely on measurements across a broad range of atomic numbers Z and mass numbers A. Light nuclei play a crucial role as benchmark systems, since nuclear-structure uncertainties (including finite-size effects, density distributions, and shell structure) are small and more testable. Measurements in boron isotopes, therefore, provide valuable constraints, significantly reducing extrapolation uncertainties, particularly in modeling K$^{-}$ interactions in neutron-rich matter relevant to astrophysical environments. These measurements provide a new benchmark for testing kaon–nucleus interaction models in the light-nuclei regime.

Section \ref{sec2:SIDDHARTA-2 kaonic boron  measurement} describes the SIDDHARTA-2 kaonic boron measurement, while Section \ref{sec3:Data selection} presents the data selection procedure. Section \ref{sec4:Results and discussion} is dedicated to the results and discussions related to the kaonic boron X-ray transition energies  and yields. Section \ref{sec5:Conclusions} presents the conclusions.

\section{The SIDDHARTA-2 kaonic boron  measurement}
\label{sec2:SIDDHARTA-2 kaonic boron  measurement}

The SIDDHARTA-2 experiment was developed to perform high-precision X-ray spectroscopy of kaonic atoms in the high-radiation environment of  the DA$\Phi$NE collider of the INFN National Laboratories in Frascati (Italy).

DA$\Phi$NE \cite{Milardi:2018sih,Milardi:2021khj,Milardi:2024efr} is an electron–positron collider operating at the center-of-mass energy corresponding to the $\phi$ resonance (1.02 GeV), which decays,  with a branching ratio of about 48.9$\%$, into a pair of charged kaons. A unique feature of the DA$\Phi$NE collider is the production of low-momentum kaons, which are emitted nearly monoenergetically at low momentum ($\Delta$p $\simeq$ 0.1$\%$), with a momentum of $\sim$ 127~MeV/c, being the ideal facility for the production and study of kaonic atoms.

\begin{figure}[htbp]
        \centering
        \includegraphics[width=0.7\textwidth]{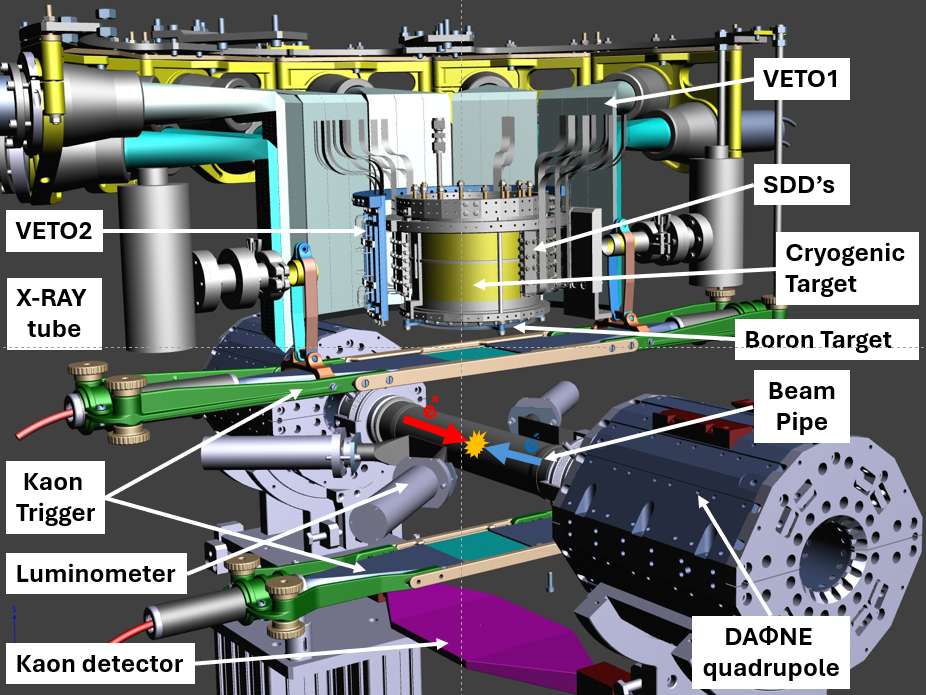}
        \caption{Schematic layout of the SIDDHARTA-2 experimental apparatus installed at the DA$\Phi$NE interaction point. The main elements, such as the kaon trigger, the cryogenic target cell surrounded by SDDs, and the X-ray tubes for detector calibration, are highlighted. The boron target used in this measurement is also indicated.}
        \label{fig:setup}
\end{figure}

The  experimental setup  (Figure \ref{fig:setup}) was installed above the DA$\Phi$NE interaction point in 2022 and took data until summer of 2024.

The main components of the experimental setup include the beam pipe, the kaon trigger system, a Mylar degrader, the luminosity monitor, a cylindrical vacuum chamber that hosts the cryogenic target, and the X-ray Silicon Drift Detectors (SDDs)  array, all enclosed by  veto systems.
The core of the apparatus is a cryogenic target system surrounded by 384 SDDs, providing a total active area of 245.8 cm$^{2}$. The target can be filled with different types of gases \cite{Sirghi:2023wok}.


The SDDs have been developed by the Fondazione Bruno Kessler (FBK) in collaboration with INFN-LNF, Politecnico of Milano and the Stefan Meyer Institute (SMI), specifically for performing kaonic
atom measurements. The excellent  energy resolution of these detectors,  FWHM of 157.8 $\pm $ 0.3 eV at 6.4 keV  \cite{Miliucci:2021wbj}, together with their excellent timing resolution of 500 ns \cite{Miliucci:2022lvn}, are fundamental for the background reduction and, consequently, the success of the measurement. The energy calibration of the SDDs is crucial to guarantee the accuracy of the kaonic atoms measurements. It is performed using a system composed of two X-ray tubes and a multi-element target made of high-purity titanium, iron, and copper strips. The X-ray tubes induce
the fluorescence emission in the target elements, and their characteristic lines are used to
calibrate the SDDs. This procedure ensures an energy calibration accuracy of a few eV within the energy range $\leq$~20~keV \cite{Sgaramella:2022rbl}, matching the requirements for precision X-ray spectroscopy.

Two main classes of background are taken into account: electromagnetic and hadronic. The electromagnetic background, which is uncorrelated with kaon production, at the time scale of SDDs response (hundreds of ns), originates from particles lost from the  DA$\Phi$NE  circulating beams as a consequence of beam–gas interactions and the Touschek effect.
The kaon trigger (KT) system, composed of two plastic scintillators positioned above and below the interaction region, is employed to detect back-to-back emitted K$^{+}$K$^{-}$  pairs. The coincidence of the signals from the two scintillators defines the trigger and enables the rejection of SDD hits that are not synchronous with kaon production.

The hadronic background arises from K$^{-}$ nuclear absorption and  K$^{+}$ decays, producing minimum ionizing particles (MIPs), primarily pions and muons, which generate signals in the SDDs correlated with the kaon trigger. To suppress this background, dedicated veto systems \cite{Bazzi:2013kwa}, located behind the SDDs and around the vacuum chamber, are employed to identify and reject MIP-induced signals. A luminosity monitor \cite{Skurzok:2020phi} placed in the longitudinal plane in front of the interaction region enables real-time background monitoring and precise measurement of the collider luminosity.   Further details on the SIDDHARTA-2 experimental setup can be found in \cite{Sirghi:2023wok}.

The kaonic boron measurement was performed in Spring 2024; a solid boron target, 2~mm thick (75$\times$ 50 mm$^{2}$)  was placed on the top scintillator of  the kaon trigger, below the vacuum chamber entrance, where the kaons were stopped to form the kaonic boron.  The data were collected over approximately one week, for a total integrated luminosity of 22~pb$^{-1}$.

\section{Data selection}
\label{sec3:Data selection}

\begin{figure}[htbp]
\centering
\mbox{\includegraphics[width=16 cm]{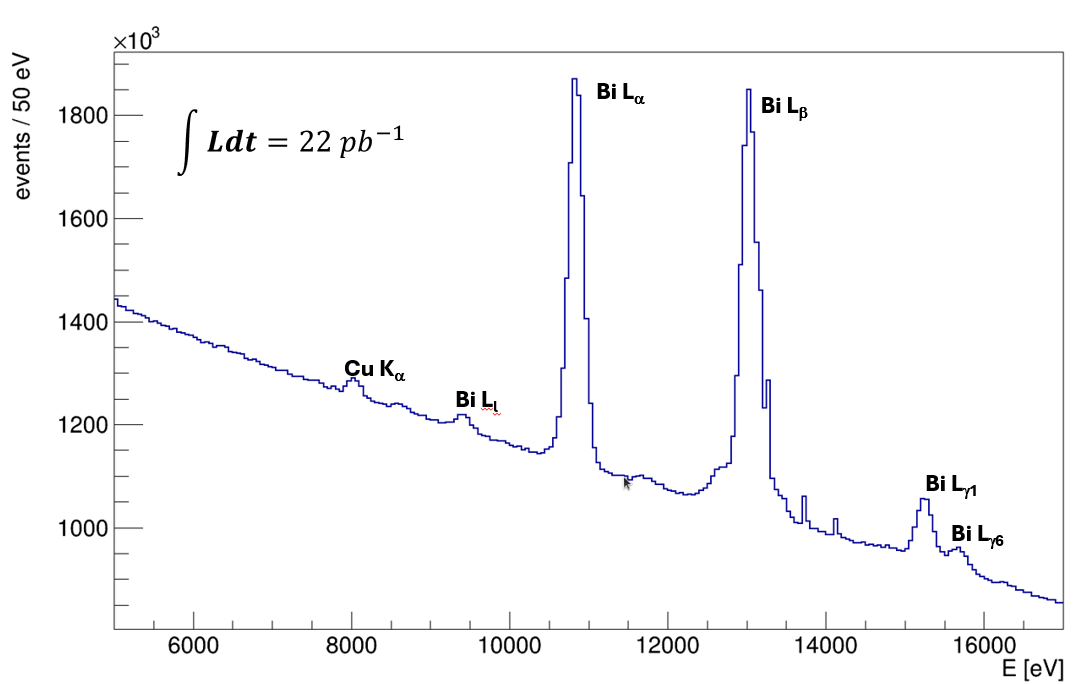}}
\caption{Raw kaonic boron energy spectrum with the corresponding  transitions: copper lines originate from components inside the vacuum chamber, while bismuth lines arise from alumina ceramic boards behind the SDDs.}
\label{fig_datanotrig}
\end{figure}

Figure \ref{fig_datanotrig} shows the raw X-ray energy spectrum acquired by the SDDs during the boron data-taking period. The spectrum exhibits several fluorescence peaks associated with X-ray emission from materials surrounding the SDDs. The  copper lines originate from electronic components of the experimental setup located inside the vacuum chamber, while the bismuth lines arise from the alumina ceramic boards positioned behind the SDDs.\\

The presence of a large continuous background prevents the direct observation of the kaonic boron signal. To suppress the large continuous background, event selection was performed using the kaon trigger (KT). Only events occurring within an 8~$\mu$s time window in coincidence with a trigger signal are selected, thereby rejecting a substantial fraction of the background. The width of the time window was optimized to ensure proper signal processing and acquisition by the front-end electronics \cite{Sirghi:2023wok}.

\begin{figure}[htbp]
\centering
\mbox{
\includegraphics[width=8.5 cm]{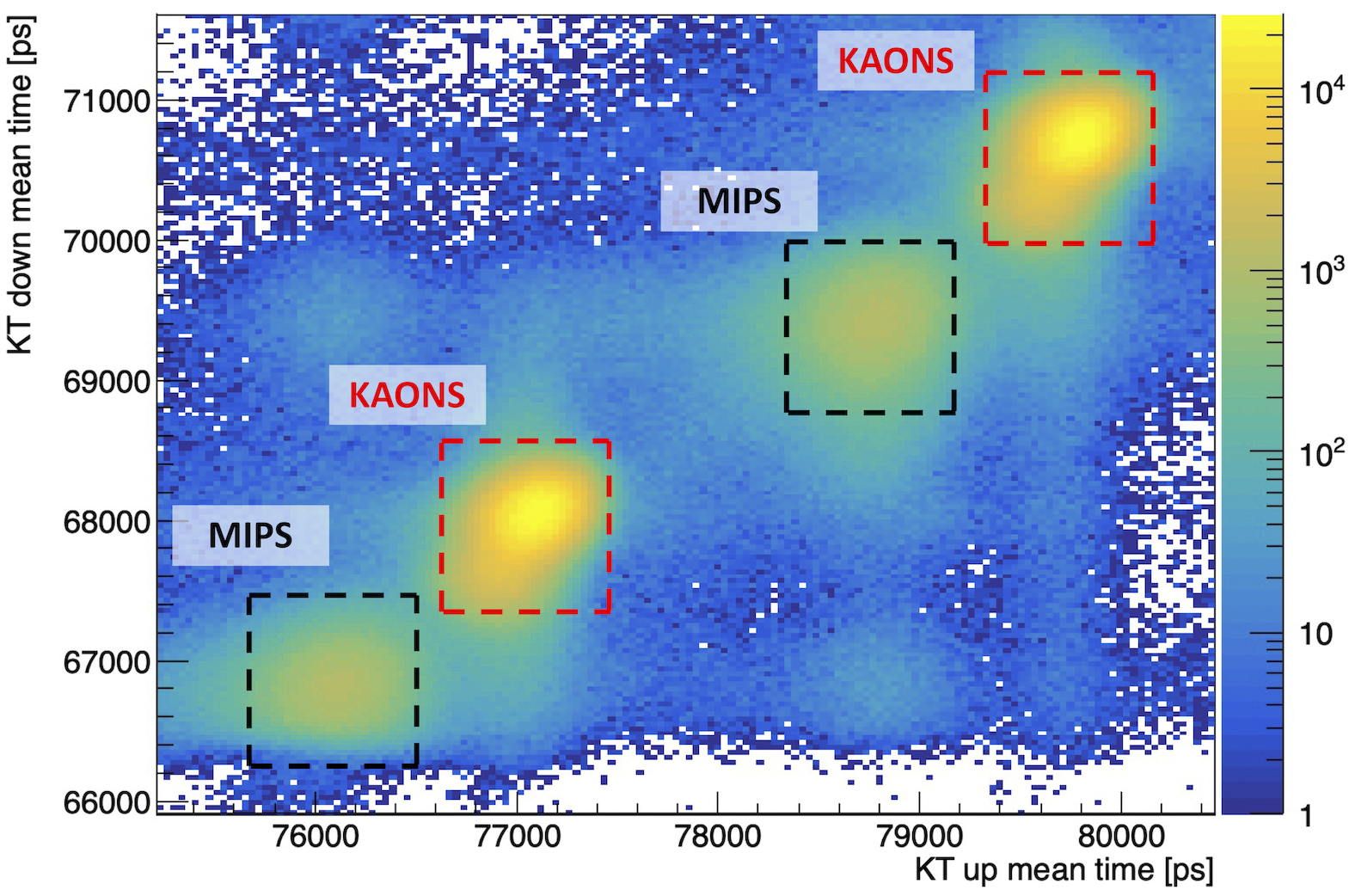}
\includegraphics[width=7.5 cm]{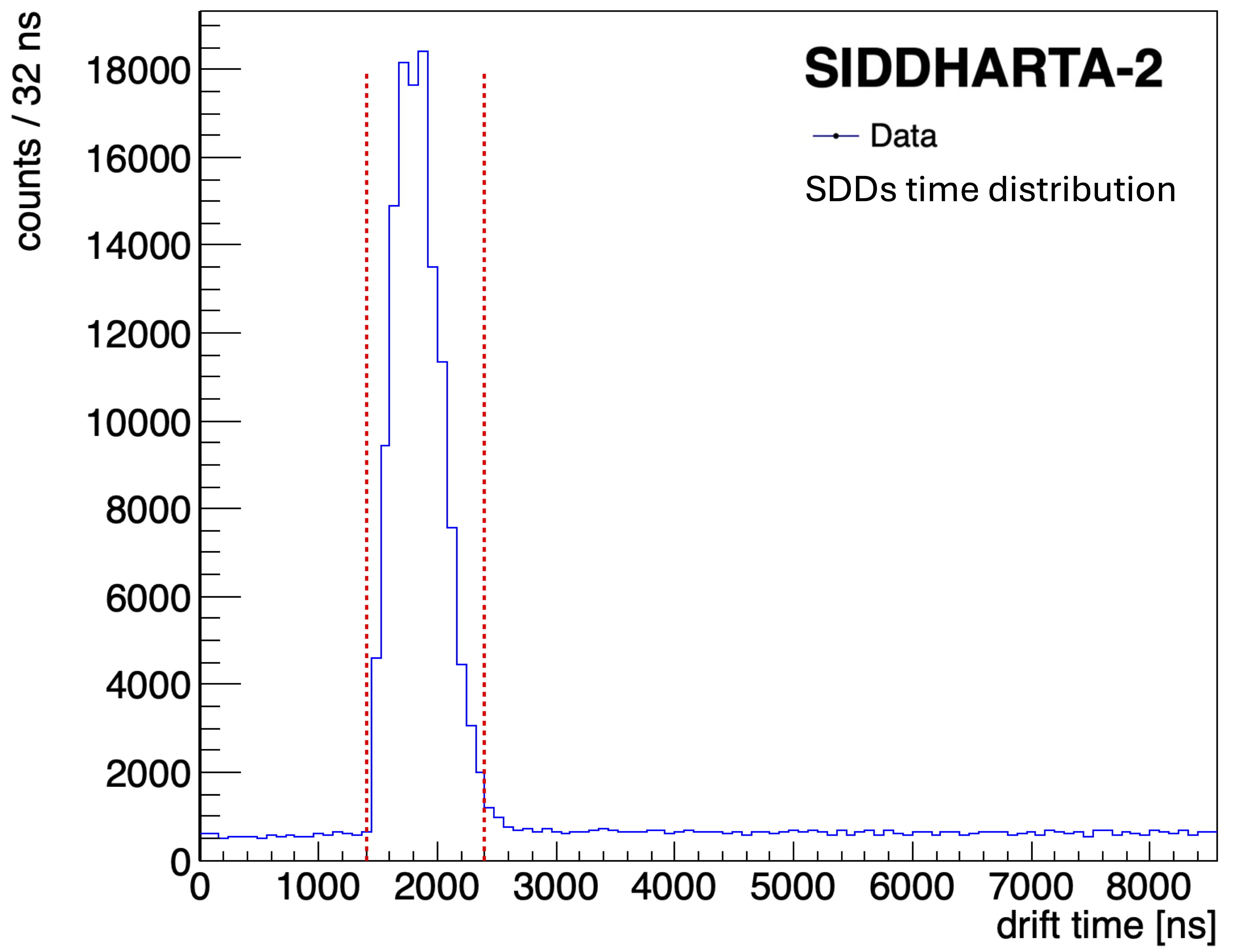}
}
\caption{Left: Two-dimensional scatter plot of KT time distributions. The coincidence events related to kaons (high intensity) are clearly distinguishable from MIPs (low intensity). Right: Time difference between the KT signals and X-ray hits on the SDDs. The dashed lines represent the acceptance window.}
\label{fig_kaon_drift}
\end{figure}

Nevertheless, minimum ionizing particles (MIPs) produced by beam–beam and beam–gas interactions can occasionally generate trigger signals when simultaneously passing through the KT scintillators. To discriminate between these MIP-induced triggers and those originating from genuine K$^{+}$K$^{-}$ pairs, a Time-of-Flight (TOF) analysis is applied \cite{Sirghi:2023wok}. This method is based on measuring the time difference between the trigger signal and the DA$\Phi$NE radio-frequency (RF), which provides a reference for the collision time. Figure \ref{fig_kaon_drift}-left shows the correlation of the mean time distribution measured by the two scintillators of the kaon trigger during the kaonic boron run, demonstrating the effectiveness of the selection cut in distinguishing K$^{+}$K$^{-}$ pairs from MIPs.
To further improve background suppression, the time difference between the KT signal and the X-ray detection time in the SDDs was used \cite{Sirghi:2023wok, Sgaramella:2023orc}, reducing the initial 8$\mu$s to 500 ns. The corresponding time distribution  shown in Figure \ref{fig_kaon_drift}-right exhibits a pronounced peak within the red dashed lines, corresponding to SDD hits in coincidence with the trigger, superimposed on a flat distribution arising from uncorrelated events. The combined use of the timing information from the KT and the SDDs allows a background reduction of approximately five orders of magnitude.

\begin{figure}[ht]
\centering
\mbox{\includegraphics[width=16.5 cm]{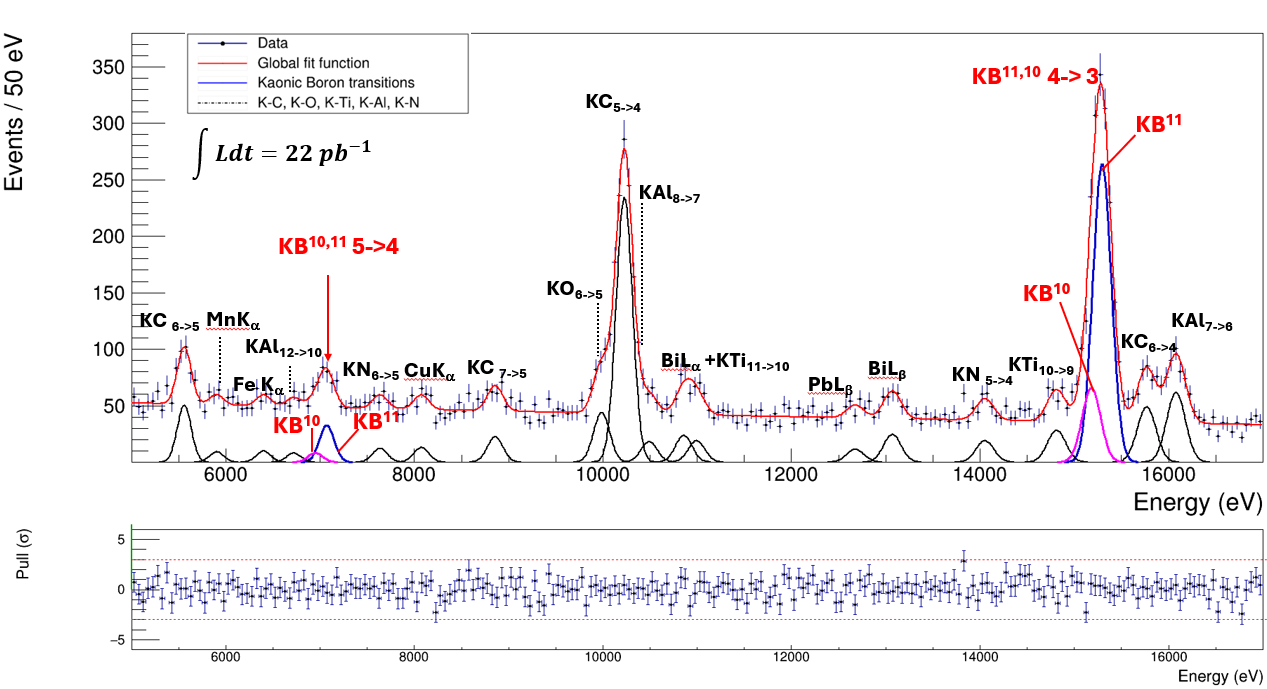}}
\caption{Top: X-ray energy spectrum and corresponding fit to the data after the background suppression procedure (see text). The kaonic boron $5g\rightarrow f4$ and $4f\rightarrow3d$ transitions are indicated in magenta for the  kaonic ${}^{10}$B and in blue for kaonic ${}^{11}$B. The fit includes all identified kaonic atom contributions and background components. The pull plot (bottom panel) shows the residuals normalized to their corresponding statistical uncertainties. }
\label{fig_fit}
\end{figure}

Figure \ref{fig_fit} shows the X-ray spectrum obtained after the implementation of the event selection. Clear signals from kaonic atoms are observed, with highlighted peaks corresponding to X-ray emissions originating from kaonic atoms formed within the boron. The other lines are due to kaons stopped in the Kapton (C$_{22}$H$_{10}$O$_5$N$_2$) entrance window and the aluminium frame of the target cell, with a strong suppression of the Bi lines (see Fig.\ref{fig_datanotrig}).

To identify the kaonic boron peaks, their transition energies were calculated using the Multiconfiguration Dirac–Fock General Matrix Element (\textsc{mcdfgme}) \cite{desclaux1975,indelicato1990}. The circular transitions  5g $\to$ 4f and 4f$\to$ 3d transitions for the two boron isotopes are listed in Table \ref{tab:KB}.

\section{Results and discussion}
\label{sec4:Results and discussion}

\subsection{Kaonic boron  transitions}

After the selection procedure, presented in Section \ref{sec3:Data selection}, the kaonic boron X-rays  lines are clearly visible (Figure \ref{fig_fit}), specifically the $5g\rightarrow4f$ and$4f\rightarrow3d$ transitions.  In addition, X-ray lines corresponding to kaonic carbon, oxygen, nitrogen, titanium and aluminium, produced by kaons stopping in the apparatus support frame and in the Kapton window of the target cell, are also observed. A detailed analysis of these lines is reported in \cite{Sgaramella:2023orc}.

The energy associated with each kaonic atom transition, reported in Figure \ref{fig_fit}, was determined through spectral fitting procedure. The energy response of the SDDs for each transition line is described by the convolution of a Gaussian with an exponential tail function to account for incomplete charge collection and electron-hole recombination effects \cite{Campbell:1990ye,CAMPBELL1997297,Gysel:2003}. 

The continuous background was modeled using a first-order polynomial combined with an exponential function. The fit was performed over an energy range from 5.5 keV to 17 keV in order to include all the observed kaonic boron lines. Since boron occurs naturally as a mixture of two stable isotopes,  80.1$\%$ for ${}^{11}$B and 19.9$\%$ for ${}^{10}$B, both contributions were included in the fit of the kaonic boron peaks, with their relative amplitudes constrained to the natural isotopic abundance ratio, $f_{10}/f_{11} = 0.199/0.801 = 0.2484 \pm 0.0047$. The kaonic boron $5g\rightarrow4f$ and $4f\rightarrow3d$ transitions were measured, and the corresponding energy values are reported in Table \ref{tab:KB}. All measured transition energies are consistent with the MCDFGME calculations within the quoted uncertainties.

\begin{table}[htbp]
\centering
\caption{Measured and calculated transition energies for kaonic boron. 
Transitions $5g \to 4f$ and $4f \to 3d$ are reported for both isotopes. 
Experimental values are given as $\mathbf{E^{(\mathrm{exp})} \pm \mathrm{(stat.)} \pm \mathrm{(sys.)}\,(eV)}$, 
while calculated values $\mathbf{E^{(\mathrm{calc})}\,(eV)}$ are obtained using the \textsc{mcdfgme} code.}
\begin{tabular}{ccc}
\hline
\textbf{Transition} & $\mathbf{E^{(\mathrm{exp})} \pm \mathrm{(stat.)} \pm \mathrm{(sys.)}\,(eV)}$ & $\mathbf{E^{(\mathrm{calc})}\,(eV)}$ \\
\hline
Kaonic ${}^{11}$B ($5g \to 4f$) & $7064.62 \pm 16.93 \pm 2.00$ & 7065.09 \\
Kaonic ${}^{10}$B ($5g \to 4f$) & $6920.96 \pm 58.23 \pm 2.00$ & 7032.87 \\
Kaonic ${}^{11}$B ($4f \to 3d$) & $15293.33 \pm 4.80 \pm 5.30$ & 15284.23 \\
Kaonic ${}^{10}$B ($4f \to 3d$) & $15180.11 \pm 20.86 \pm 5.30$ & 15214.48 \\
\hline
\end{tabular}
\label{tab:KB}
\end{table}

The systematic uncertainty of 2.0 eV associated with the kaonic boron $5g\rightarrow4f$ transition is mainly defined by the energy calibration accuracy in the 3.0–12.0 keV range \cite{Curceanu:2023yuy}. The systematic uncertainty of 5.3 eV for the kaonic boron  $4f\rightarrow3d$  transition was estimated from the residual of the fit to the Bi L$_{\gamma1}$ peak at 15247.7~eV in the non-triggered spectrum (Fig. \ref{fig_datanotrig}), which lies at an energy close to that of the kaonic boron  $4f\rightarrow3d$ transition.

These measurements provide new data for the database of  light kaonic atoms from solid targets, enabling investigations of kaon interactions with few-nucleon systems.
 Our precision measurement of the kaonic-boron $4f\rightarrow3d$ X-ray transitions in  ${}^{11}$B reveals no statistically significant deviation from purely electromagnetic (QED) predictions. Interpreting these results in terms of phenomenological kaon–nucleus optical potentials, and, within model-dependent assumptions, in terms of the complex kaon–nucleus scattering amplitude \cite{Friedman2007,Friedman_2012,Weise2008,RamosOset2000}, can place stringent constraints and disfavor potentials producing large energy shifts or widths in boron.

\subsection{Kaonic Boron yields}

The X-ray yield of a given transition is defined as the probability for the kaonic atom to de-excite via X-ray emission corresponding to that transition. Measuring the yields of different transitions is essential to probe the atomic cascade, where X-ray emission competes with Auger processes, Coulomb de-excitation, and for low-n states, with nuclear absorption of the kaon. The lack of experimental data has so far limited the development of fully defined theoretical cascade models needed to constrain their free parameters.
\par The Monte Carlo simulations were carried out with the GEANT4 toolkit, implementing a realistic description of the experimental geometry and materials and accurately reproducing the data-taking conditions, including trigger and SDD detection efficiencies \cite{Sirghi:2023scw}. The simulations were employed to determine the fraction of kaons stopping in the boron target, a key input for the extraction of the absolute X-ray yields. 

The absolute yield (Y)  of a transition is defined as the ratio of the experimental detection ($\epsilon^{exp}$) to the Monte Carlo efficiency ($\epsilon^{MC}$).

The $\epsilon^{exp}$ is obtained by normalizing the number of measured X-rays ($N_{X-ray}^{exp}$) to the number of kaon triggers ($N_{KT}^{exp}$) and the active area of the detectors. Similarly, the $\epsilon^{MC}$ is defined as the ratio  of the number of simulated detected X-rays ($N_{X-ray}^{MC}$), to the number of simulated  kaon triggers ($N_{KT}^{MC}$) and the active area of the detectors. The kaonic atom X-rays were generated at the  $K^{-}$  stopping positions, and were isotropically emitted with a 100$\%$ yield for each transition and each kaonic atom.

The absolute yield Y of the  X-ray transition for the kaonic boron is then given by:

\begin{equation}
\centering
Y = \frac{\epsilon^{exp}}{\epsilon^{MC}} = \frac{N_{X-ray}^{exp}/N_{KT}^{exp}}{N_{X-ray}^{MC}/N_{KT}^{MC}}
\label{eq:yield}
\end{equation}

\noindent

The transition yields are extracted from data corresponding to the integrated luminosity of 22~pb$^{-1}$, collected under stable running conditions, ensuring a reliable comparison with Monte Carlo simulations.

The number of events for  kaonic boron transitions was obtained from a fit to the energy spectrum (Fig.~\ref{fig_fit}) and was used to determine the X-ray yields of the $5g\rightarrow4f$ and $4f\rightarrow 3d$ transitions. The extracted event counts, with statistical uncertainties from the fit, are summarized in Table~\ref{tab:KB_yields}. The corresponding absolute yields are derived using Eq.~(\ref{eq:yield}) and reported in Table~\ref{tab:KB_yields}, including statistical and systematic uncertainties. 

\begin{table}[htbp]
        \centering
        \caption{Event counts for kaonic boron transitions extracted from the fit to the spectrum in Fig.~\ref{fig_fit}, together with the corresponding absolute yields for the $5g\rightarrow4f$ and $4f\rightarrow3d$ transitions.}
        \begin{tabular}{ccc}
            \hline
            \textbf{Transition} &   \textbf{Number of events}   &   \textbf{Measured Yield per Kaon Stopped}  \\
            \hline
            Kaonic ${}^{11}$B ($5g \to 4f$) &  127 $ \pm 21$   & $0.076 \pm 0.013~(stat) ^{+0.012}_{-0.011}~(sys.)$   \\
            Kaonic ${}^{10}$B ($5g \to 4f$) &  32 $ \pm 5$   & $0.079 \pm 0.014~(stat) ^{+0.013}_{-0.011}~(sys.)$   \\
             Kaonic ${}^{11}$B ($4f \to 3d$) &  1301 $ \pm 64$   &  $0.115 \pm 0.006~(stat) ^{+0.002}_{-0.005}~(sys.)$   \\
            Kaonic ${}^{10}$B ($4f \to 3d$) &  323 $ \pm 16$   &  $0.107 \pm 0.007~(stat) ^{+0.002}_{-0.005}~(sys.)$   \\
          
            \hline
        \end{tabular}
        \label{tab:KB_yields}
    \end{table}

Two main sources of systematic uncertainty were considered. The first arises from the experimental uncertainty on the thickness of the detector windows crossed by the X-rays before reaching the silicon detectors. This contribution was estimated by varying the window thicknesses in the Monte Carlo simulation within their experimental tolerances. All window thicknesses were varied by $\pm 5~\mu m$, while one window with larger uncertainty was varied by $\pm$ 20\% of its nominal value. 
The second and dominant contribution originates from the experimental uncertainty on the thickness of the boron foils forming the target, which directly affects the kaon stopping probability in the boron target and therefore the expected X-ray production rate in the Monte Carlo simulation. This uncertainty was evaluated by varying the boron foil thickness of 2~mm  by  $\pm$ 0.2~mm  in the simulation. This contribution corresponds to a relative uncertainty of the extracted yield  of +15.7~\%  and -13.6~\%for the $5g\rightarrow4f$ transition and $\pm$ 1.8\% for the  $4f\rightarrow3d$ transition.

The total systematic uncertainty was obtained by combining the individual contributions in quadrature. Since the variations of the Monte Carlo parameters lead to non-linear effects on the extracted yield, the upward and downward deviations from the nominal value are different, and the systematic uncertainties are therefore reported asymmetrically.

The results reported in Table~\ref{tab:KB_yields} constitute the first experimental determination of the $5g\rightarrow4f$ and $4f\rightarrow3d$ transition yields, providing new data to refine cascade models for kaonic boron and, more generally, for kaonic atoms.

\section{Conclusions}
\label{sec5:Conclusions}

A precision measurement of X-ray transitions in kaonic boron has been performed by the SIDDHARTA-2 collaboration at the DA$\Phi$NE collider. The energies and yields of the $5g\rightarrow4f$ and $4f\rightarrow3d$ transitions were determined for both boron isotopes, $^{10}$B and $^{11}$B. The precision measurement of the $4f\rightarrow3d$ transitions in kaonic $^{11}$B shows no statistically significant deviation from electromagnetic (QED) expectations. 
By translating these limits into bounds on phenomenological kaon–nucleus optical-potential parameters, and, in a model-dependent way, on the complex scattering amplitude, we constrain and disfavor scenarios predicting large shifts or widths in boron. These results demonstrate the feasibility of studying light kaonic atoms using solid targets, opening the way for future investigations of kaon interactions with few-nucleon systems. This establishes kaonic boron as a key system for future precision studies of strong-interaction effects in light nuclei.

\section*{Acknowledgments}
We thank C. Capoccia from LNF-INFN and H. Schneider, L. Stohwasser, and D. Pristauz-Telsnigg from Stefan Meyer-Institut for their fundamental contribution in designing and building the SIDDHARTA-2 setup. We gratefully acknowledge the Polish high-performance computing infrastructure PLGrid (HPC
Center: ACK Cyfronet AGH) for providing computer
facilities and support within the computational grant
no. PLG/2025/018524. We also thank INFN-LNF and
the DA$\Phi$NE staff for the excellent working conditions
and their ongoing support. Special thanks to Catia
Milardi for her continued support and contribution
during the data taking. Part of this work was supported by the INFN (KAONNIS project); the Austrian
Science Fund (FWF): [P24756-N20 and P33037-N];
the Croatian Science Foundation under the project
IP-2022-10-3878; the EU STRONG-2020 project (Grant
Agreement No. 824093); the EU Horizon 2020 project
under the MSCA (Grant Agreement 754496); the
Japan Society for the Promotion of Science JSPS
KAKENHI Grant No. JP18H05402, JP22H04917;
the Polish Ministry of Science and Higher Educa-
tion grant No. 7150/E-338/M/2018 and the Polish
National Agency for Academic Exchange (grant no
PPN/BIT/2021/1/00037). This article/publication is
based upon work from COST Action CA24131 (ENRICH),
supported by COST (European Cooperation in Science
and Technology, http://www.cost.eu/).

\section*{References}
\bibliography{iopart-num}

\end{document}